\documentclass[aps,twocolumn,showpacs,eqsecnum,amsmath,amssymb]{revtex4}
\usepackage{graphicx}
\usepackage{dcolumn}
\usepackage{bm}
\begin{document}
\title{Two-field Warm Inflation and Its Scalar Perturbations on Large Scales}
\author{Yang-Yang Wang}
\email{wangyy@mail.bnu.edu.cn}
\affiliation{Department of Physics, Beijing Normal University, Beijing 100875, China}
\author{Xiao-Min Zhang}
 \email{zhangxm@mail.bnu.edu.cn}
  \affiliation{School of Science, Qingdao University of Technology, Qingdao 266033, China}
\author{Jian-Yang Zhu}
\thanks{Corresponding author}
 \email{zhujy@bnu.edu.cn}
\affiliation{Department of Physics, Beijing Normal University, Beijing 100875, China}
\begin{abstract}
We explore the homogeneous background dynamics and the evolution of generated perturbations of cosmological inflation that is driven by multiple scalar fields interacting with a perfect fluid.
Then we apply the method to warm inflation driven by two scalar fields and a radiation fluid, and present general results about the evolution of the inflaton and radiation. After decomposing the perturbations into adiabatic and entropy modes, we give the equation of motion of adiabatic and entropy perturbations on large scales. Then, we give numerical results of background and perturbation equations in a concrete model (the dissipative coefficient $\Gamma \propto H$). At last, we use the most recent observational data to constrain our models and give the observationally allowed regions of parameters. This work is a natural extension of warm inflation to multi-field cases.
\end{abstract}
\maketitle

\section{Introduction}

Inflation has become one of the central paradigm in modern cosmology, because it solves many problems of standard cosmology and provides an origin of large-scale structure \cite{the inflationary universe,a new inflationary}. In inflation theory, a most common model is that cosmological inflation is driven by a scalar field whose potential dominates other forms of energy density. In standard inflation, cosmological expansion and reheating are two
distinguished periods and we still know little about the reheating
process. Warm inflation is an important inflationary model and it combine the cosmological expansion and the production of the radiation into one process, so the universe can become radiation-dominated smoothly \cite{warm inflation and}. In warm inflation, dissipative effects are important during the inflation period, so that radiation production occurs concurrently with cosmological expansion. Besides, recent observations imply that chaotic inflation with monomial quadratic potential and natural inflation are now disfavored for predicting too large tensor-to-scalar ratio $r$. In warm inflation, curvature perturbation are dominated by thermal fluctuation which is usually much stronger than quantum fluctuation, while tensor perturbation remain the same to the cold inflation results. Therefore, many inflationary models are in accordance with the observational data again for a decreased $r$ in warm regime.

A different possible way to generate perturbations in agreement with observations is so-called multi-field inflationary model. Although single field inflation may seems appealing from the perspective of simplicity and economy, the microphysical origin of inflation still remains unclear and there is not theoretical reason to expect only one field to be important in the early Universe. In fact, fundamental physics, such as sting theory, commonly predicts the existence of multiple scalar fields \cite{curvature and isocurvature,cosmology}.

Compared to single-field inflation, a key feature of multi-field inflation is a relatively large non-Gaussianity. However, now the observations of non-Gaussianity is not precise enough to distinguish between inflationary models. Therefore it is important to study the effects of multi-field inflation and how they are constrained by observational data. In addition, the content of the Universe is commonly assumed to be a mixture of fluids and scalar fields, and there has been increasing interests focused on multi-component cosmology. In this paper we investigate cosmological inflation driven by multiple scalar fields and an interacting perfect fluid, and then apply the formalism to warm inflation in a two-field case. This work is a natural extension of warm inflation to multi-field cases.

This paper is organized as follows. In Sec. II we introduce the governing background and perturbation equations of multiple scalar fields interacting with a perfect fluid. In Sec. III we apply the formalism to warm inflation and obtain the evolution equations of curvature and entropy perturbation in a two-field case. In Sec. IV, we give numerical results in a representative case with the dissipative coefficient $\Gamma \propto H$, then we use the most up-to-data observational data to constrain our models and give the observationally allowed region of parameters. To conclude, we present some summaries and comments in Sec. IV. In this paper, we redefined some slow-roll parameters of the inflation and treat the radiation as a perfect fluid.

\section{\label{se2}Multi-component inflation}

Let us study the inflation in homogeneous and isotropic background. We consider a spatially flat Friedmann-Robertson-Walker (FRW) metric of the form \begin{equation}
ds^2=-dt^2+a(t)^2\delta_{ij}dx^idx^j,
\end{equation}
where $a(t)$ is the scale factor, and $t$ is cosmic time. We use Planck Unit:$$8\pi G=k_B=\hbar=c=1,$$ where $G$ is Newton's gravitational constant, $k_B$ is Boltzmann's constant,
$\hbar$ is the reduced Planck's constant, and $c$ is the speed of light. In this work, Greek indices $\mu$, $\nu$, $\lambda$ denote spacetime dimensions, and Latin indices $I$, $J$, $\mathcal{N}$ denote different scalar fields. Repeated spacetime indices are summed over.

First we consider a $\mathcal{N}$-fields model with Lagrangian density \cite{non-gaussianities in two-field}:
\begin{equation}
\mathcal{L}= - \frac{1}{2} \Sigma_I^\mathcal{N} g^{\mu \nu} \nabla_{\mu} \varphi_I
\nabla_{\nu} \varphi_I - V (\varphi), \label{1}
\end{equation}
which is minimal coupled to gravity, where $V(\varphi)=V(\varphi_1,
\varphi_2, \ldots, \varphi_\mathcal{N})$. We assume that there exists a perfect fluid
and the interaction between scalar fields and the fluid causes a
phenomenological dissipative term $\Gamma \dot{\varphi_I}$ in the equation of
motion. In general, $\Gamma = \Gamma (\varphi_1, \varphi_2, \ldots, \varphi_\mathcal{N},
\rho_f)$, $\rho_f$ is the energy density of the perfect fluid. From Eq. \eqref{1} we can
get the equation of motion in the presence of a perfect fluid:
\begin{equation}
\ddot{\varphi_I} + (3 H + \Gamma) \dot{\varphi_I} + V_{\varphi_I} = 0,
\end{equation}
where $H\equiv \dot{a}/a$ is the Hubble parameter, $V_{\varphi_I} = \frac{\partial V}{\partial \varphi_I}$, overdots represent derivatives with respect to cosmic time. In a spatially flat FRW
universe, $H$ is determined by:
\begin{equation}
H^2 = \frac{8 \pi G}{3} \left( \frac{1}{2} \Sigma_I^\mathcal{N} \dot{\varphi_I}^2 + V (\varphi) + \rho_f \right),
\end{equation}
and the continuity equation of the perfect fluid
\begin{equation}
 \dot{\rho_f} + 3 H (1 + \omega) \rho_f = \Gamma \Sigma_I \dot{\varphi_I}^2,
\end{equation}
where $\omega = p_f/\rho_f$, $p_f$ is the pressure of the fluid.

As in single-field inflation, we define some slow-roll parameters of the background quantities,
\begin{gather}
\epsilon = - \frac{\dot{H}}{H^2} = \Sigma_I \epsilon_I + \epsilon_f, \quad
\epsilon_I = \frac{1}{2} \frac{\dot{\varphi}_I^2}{H^2}, \quad
\epsilon_f = \frac{2}{3} \frac{\rho_f}{H^2}, \nonumber \\
\eta = - \frac{\ddot{H}}{2 H \dot{H}} = \Sigma_I \frac{\epsilon_I}{\epsilon} \eta_I + \frac{\epsilon_f}{\epsilon} \eta_f, \\
\eta_I = - \frac{\ddot{\varphi_I}}{H \dot{\varphi_I}}, \quad
\eta_f = - \frac{1}{2} \frac{\dot{\rho}_f}{H \rho_f}. \nonumber
\end{gather}

The slow-roll conditions are $\epsilon<1$, $\left|\eta\right|<1$. In slow-roll approximation, we have
\begin{gather}
(3 H + \Gamma) \dot{\varphi_I} + V_{\varphi_I} = 0, \label{sl1}\\
3 H (1 + \omega) \rho_f = \Gamma \Sigma_I \dot{\varphi_I}^2. \label{sl2}
\end{gather}
In order to study the evolution of the linear perturbations, we decompose
each of the scalar fields into a spatially homogenous background field and its
fluctuations $\varphi_I (x, t) \rightarrow \varphi_I (t) + \delta \varphi_I
(x, t)$. The line element of the FRW metric can be written as
\begin{multline}
d s^2 = - (1 + 2 A) d t^2 + 2 a \partial_i B d x^i d t \\
+ a^2 ((1 - 2 \psi) \delta_{ij} + 2 \partial_i \partial_j E) d x^i d x^j,
\end{multline}
and the gauge-invariant comoving curvature perturbation is given by \cite{oscillatory power spectrum,adiabatic and entropy2}
\begin{equation}
\mathcal{R} = \psi - H \frac{\delta q}{p + \rho},
\end{equation}
where $\delta q = \Sigma_I \delta q_I + \delta q_f$ is the total momentum
density perturbation, and $p$ and $\rho$ are total pressure and energy
density \cite{cosmological perturbations in}. The momentum perturbations of each components are given by
\begin{eqnarray}
\delta q_I &=& - \dot{\varphi_I} \delta \varphi_I, \\
\delta q_f &=& a (p_f + \rho_f) (B + \delta u),
\end{eqnarray}
where $\delta u$ is the scalar velocity potential of the fluid, and from above
definition we know $\delta u = \frac{\delta q_f}{a (p_f + \rho_f)} - B$. The
four-velocity of the fluid is defined by
\begin{eqnarray}
u^\mu &=& \frac{1}{a} (1 - A, \partial^i \delta u), \\
u_\mu &=& a (- 1 - A, \partial_i \delta u + \partial_i B).
\end{eqnarray}
The variation of the scalar field's equation of motion leads to:
\begin{multline}
\ddot{\delta \varphi_I} + (3 H + \Gamma) \dot{\delta \varphi_I} +
\frac{k^2}{a^2} \delta \varphi_I + \Sigma_J V_{\varphi_I \varphi_J} \delta
\varphi_J + \dot{\varphi_I} \delta \Gamma \\
= - (2 V_{\varphi_I} + \Gamma \dot{\varphi_I}) A + \dot{\varphi_I} \left( \dot{A} + 3 \dot{\psi} + \frac{k^2}{a^2} (a^2 \dot{E} - a B) \right), \label{phi}
\end{multline}
and the perturbation of energy and momentum conservation equation of the fluid is given by \cite{xpand: an algorithm}
\begin{multline}
\dot{\delta \rho_f} + 3 H (\delta p_f + \delta \rho_f) - \frac{k^2}{a^2}
\delta q_f + \frac{k^2}{a} (p_f + \rho_f) B \\
- \delta \Gamma \Sigma_I \dot{\varphi_I}^2 - 2 \Gamma \Sigma_I \dot{\varphi_I} \dot{\delta \varphi} \\
= k^2 \dot{E} (p_f + \rho_f) - \Gamma A \Sigma_I \dot{\varphi_I}^2 + 3 (p +
\rho) \dot{\psi}, \label{energy}
\end{multline}
\begin{equation}
\dot{\delta q_f} + 3 H \delta q_r + \delta p + \Gamma \Sigma_I
\dot{\varphi_I} \delta \varphi_I = - (p_f + \rho_f) A, \label{momentum}
\end{equation}
where $k$ is the wave number in Fourier space.

The Einstein equation of the multi-component system is given by
\begin{equation}
G_{\mu \nu} = T_{\mu \nu}^{(\varphi)}+T_{\mu \nu}^{( f )},
\end{equation}
where
\begin{equation}
T_{\mu \nu}^{(\varphi)} \! = \! \Sigma_I^\mathcal{N} \partial_\mu \varphi_I \partial_\nu \varphi_I - g_{\mu \nu} \! \left( \frac {1}{2} \Sigma_I^\mathcal{N} \partial^\lambda \varphi_I \partial_\lambda \varphi_I +V(\varphi) \right),
\end{equation}
\begin{equation}
T_{\mu \nu}^{(f)} \! = \! (\rho_f + p_f) u_\mu u_\nu + p_f g_{\mu \nu},
\end{equation}
where $G_{\mu \nu}$ is the Einstein tensor, and $T_{\mu \nu}^{(\varphi)}$, $T_{\mu \nu}^{(f)}$ are the energy-momentum tensor of scalar fields and perfect fluid respectively.

The perturbation equations of Einstein's field equations are:
\begin{multline}
\delta \rho_f + \Sigma_I \dot{\varphi_I} \delta \varphi_I + \Sigma_I
V_{\varphi_I} \delta \varphi_I = - 2 (V + \rho_f) A \\
+ \frac{2 k^2 H}{a^2} (a B
- a^2 \dot{E}) - \frac{k^2}{a^2} \psi - 6 H \dot{\psi}, \label{e1}
\end{multline}
\begin{equation}
\shoveleft \delta q_f - \Sigma_I \dot{\varphi_I} \delta \varphi_I = - 2 H A - 2 \dot{\psi}, \label{e2}
\end{equation}
\begin{equation}
\shoveleft a^2 (\ddot{E} + 3 H \dot{E}) - a (\dot{B} - 2 H B) - A + \psi = 0, \label{e3}
\end{equation}
\begin{multline}
\delta p_f + \Sigma_I \dot{\varphi_I} \dot{_{} \delta \varphi_I} -
V_{\varphi_I} \delta \varphi_I \\
= - \frac{k^2}{a^2} (a \dot{B} - a^2 \ddot{E})
- \frac{k^2 H}{a^2} (2 a B - a^2 \dot{E}) - \frac{k^2}{a^2} A \\
+ 2 (V - p_f) A
+ 2 H \dot{A} + \frac{k^2}{a^2} \psi + 6 H \dot{\psi} + 2 \ddot{\psi}. \label{e4}
\end{multline}
Eqs.~\eqref{e1}-\eqref{e4} are, respectively, the $G_0^0$ component of the field equation,
the $G_i^0$ component, the trace-free part of the $G_i^j$ and the $G_i^i$
component \cite{perturbations in cosmologies}.

\section{\label{se3}Application to warm inflation}

In warm inflation, the density perturbations are mainly sourced by thermal
noise \cite{non-gaussianity in fluctuations}, and metric fluctuations has little effect on small scales \cite{scalar perturbation spectra,a relativistic calculation}.
When $\frac{k}{a H} \gg 1$, inflaton fluctuations $\delta \varphi_I$ are
described by a Langevin equation \cite{warm inflation and}
\begin{equation}
\ddot{\delta \varphi_I} (k, t) + (3 H + \Gamma) \dot{\delta \varphi_I} (k, t) +
\frac{k^2}{a^2} \delta \varphi_I = \xi_I (k, t), \label{le}
\end{equation}
where $\xi_I (k, t)$ is a stochastic noise source and different components of $\xi_I (k,t)$ is independent of each other. From the equation above, we know there is no direct coupling between different components of field perturbations when dropping out metric
fluctuations on small scales. If the temperature is sufficiently high, the noise source is
Markovian \cite{scalar perturbation spectra},
\begin{equation}
\langle \xi_I (k, t) \xi_J (- k', t') \rangle = 2 \Gamma T a^{- 3} \delta_{I\!J} \delta^3 (k -
k') \delta (t - t'). \label{im}
\end{equation}

Thermal noise is transferred to inflation field mostly on small scales, and as the wavelength of perturbations expands, the thermal effects decrease until the fluctuation amplitude freezes out.

At horizon-crossing, for T-dependent dissipative coefficients the thermal fluctuations produce a power spectrum of perturbation \cite{density fluctuations from}
\begin{equation}
\mathcal{P}_{\varphi_I} = k^{- 3} \frac{\sqrt{\pi}}{2} H^{1 / 2} (3 H + \Gamma)^{1 / 2} T, \label{ip}
\end{equation}

After horizon-crossing, we have to take into account the influence of metric
perturbations \cite{cosmological inflation and}. For simplicity, we will work in spatially-flat gauge, in which
$E = \psi = 0$. Since there are only two degrees of freedom of metric
perturbation, only two of the equations of Eqs.\eqref{e1}-\eqref{e4} are independent. Working with Eq. \eqref{e1} and Eq. \eqref{e2}, we can get $A$ and $B$ in terms of other
perturbation variables by solving these two equations algebraically
\begin{equation}
A = \frac{- \delta q_r + \Sigma_I \dot{\varphi_I} \delta \varphi_I}{2 H}, \label{A}
\end{equation}
\begin{multline}
B = - \frac{a}{4 k^2 H^2} (24 H^2 \delta q_r \\
+ (4 \rho_r + \Sigma_I \dot{\varphi_I}^2) (- \delta q_r + \Sigma_I \dot{\varphi_I} \delta \varphi_I) \\
+ 2 H (3 \dot{\delta q_r} + 4 \Gamma \Sigma_I \dot{\varphi_I} \delta \varphi_I
+ \Sigma_I \ddot{\varphi_I} \delta \varphi_I - \Sigma_I \dot{\varphi_I}
\dot{\delta \varphi_I})). \label{B}
\end{multline}

In warm inflation, we usually treat the radiation as a perfect fluid, so the
above results can be applied here. For the radiation fluid, $p_r = \frac{1}{3}
\rho_r$, $\delta p_r = \frac{1}{3} \delta \rho_r$, where $p_r$ and $\rho_r$ are the pressure and energy density of the radiation, and $\delta p_r, \delta \rho_r$ are their perturbations respectively. With these relations, we can substitute Eq. \eqref{energy} into Eq. \eqref{momentum} and yield:
\begin{multline}
\ddot{\delta q_r} + 7 H \dot{\delta q_r} + 3 \left( 7 H^2 + \dot{H} +
\frac{1}{3} \frac{k^2}{a^2} \right) \delta q_f + \\
\Sigma_I \left( \frac{1}{3} \dot{\varphi_I}^2 \delta \Gamma \!+\! (4 H \Gamma \!+\! \dot{\Gamma}) \dot{\varphi_I}
\delta \varphi_I \!+\! \Gamma \ddot{\varphi_I} \delta \varphi_I \!+\! \frac{5}{3}
\Gamma \dot{\varphi_I} \dot{\delta \varphi_I} \right) \\
 = \frac{1}{3}
\frac{k^2}{a} \rho_r B - \frac{16}{3} H \rho_r A + \frac{1}{3} \Gamma \Sigma_I
\dot{\varphi_I}^2 A - \frac{4}{3} \rho_r \dot{A}. \label{fe}
\end{multline}
The relationship between energy density $\rho_r$ and temperature of radiation
is $\rho_r = \frac{\pi^2}{30} g_{\ast} T^4$, where $g_{\ast}$ is the effective
particle number of radiation. Now we define two new parameters describing the $\varphi_I$ dependence and temperature dependence of the damping term $\Gamma$($\Gamma = \Gamma (\varphi_1, \varphi_2, \ldots, \varphi_{\mathcal{N}}, \rho_r)$):
\begin{eqnarray}
\beta_I = \frac{\Gamma_{\varphi_I} V_{\varphi_I}}{\Gamma V}, \quad
c = \frac{T
\Gamma_T}{\Gamma} = \frac{4 \rho_r \Gamma_{\rho_r}}{\Gamma},
\end{eqnarray}
where $\beta_I < 1+r$, $\beta_I$ are slow-roll parameters \cite{power spectrum for}, but $c$ is not required to be small. In
order to go back to cold inflation $(\Gamma = 0)$ when $T = 0$, we require
$\Gamma_T > 0$, so $c$ is positive defined. Considering the consistency of
warm inflation \cite{density fluctuations from, consistency of the}, we set $0 < c < 4$.

Then $\Gamma_{\varphi_I} = \frac{\partial \Gamma}{\partial \varphi_I}$,
$\Gamma_{\rho_r} = \frac{\partial \Gamma}{\partial \rho_r}$ can be denoted by
$\beta_I$, $c$ and the corresponding background quantities. So we have \cite{consistency of the2}
\begin{eqnarray}
\delta \Gamma &=& \Sigma_I \Gamma_{\varphi_I} \delta \varphi_I +
\Gamma_{\rho_r} \delta \rho_r, \\
\dot{\Gamma} &=& \Sigma_I \Gamma_{\varphi_I}
\dot{\varphi_I} + \Gamma_{\rho_r} \dot{\rho_r}.
\end{eqnarray}

Substitute $\delta \Gamma, \dot{\Gamma}, A, B$ into Eqs.~\eqref{phi} and \eqref{fe}, and
keep the leading order, we find
\begin{multline}
\ddot{\delta q_r} + (7 - c) H \dot{\delta q_r} + \left( 12 - 3 c + \frac{1}{3}
\frac{k^2}{a^2 H^2} \right) H^2 \delta q_r \\
= \gamma \Sigma_I (5 H \dot{\delta q_I} + (12 - 3 c) H^2 \delta q_I), \label{qr}
\end{multline}
\begin{equation}
\ddot{\delta q_I} + 3 H (1 + \gamma) \dot{\delta q_I} + \frac{k^2}{a^2} \delta q_I
= 0. \label{qi}
\end{equation}
where $\gamma = \frac{\Gamma}{3 H}$ describes the dissipative strength in warm inflation.

We define a new variable $z = \frac{k}{a H}$, then
\begin{eqnarray}
\frac{d}{d t} &=& - \frac{k}{a} (1 - \epsilon) \frac{d}{d z}, \\
\frac{d^2}{d t^2} &=& (1 - \epsilon) \frac{k H}{a} \frac{d}{d z} + (1 - 2 \epsilon) \frac{k^2}{a^2} \frac{d^2}{d z^2}.
\end{eqnarray}

Replace time variable $t$ with $z$, and keep the leading order, Eqs.~\eqref{qr} and \eqref{qi} can be put in the form
\begin{multline}
z^2 \delta q_r'' - (6 - c) z \delta q_r' + \left( 12 - 3 c + \frac{1}{3} z^2
\right) \delta q_r \\
= \gamma \Sigma_I (- 5 z \delta q_I' + (12 - 3 c) \delta q_I), \label{qrz}
\end{multline}
\begin{equation}
z^{} \delta q_I'' - (2 + \gamma) \delta q_I' + z \delta q_I = 0, \label{qiz}
\end{equation}
where a prime denotes a derivative with respect to $z$.

From Eq. \eqref{qiz} we know that in the large-scale limit, i.e., $z = \frac{k}{a H} \rightarrow
0$, $\delta q_I$ is a constant in the slow-roll approximation. So we drop out
the $\delta q_I'$ term on the right-hand side of Eq. \eqref{qrz}.
\begin{multline}
z^2 \delta q_r'' - (6 - c) z \delta q_r' + \left( 12 - 3 c + \frac{1}{3} z^2
\right) \delta q_r \\
= \gamma \Sigma_I ((12 - 3 c) \delta q_I). \label{qrh}
\end{multline}

This is an inhomogeneous Bessel differential equation. The solution is given
by adding the homogeneous solution to a particular solution. The homogeneous
solution of \eqref{qrh} can be found in terms of Bessel functions
\begin{multline}
\delta q_r^h = C_1 z^{\frac{7 - c}{2}} J_{\tilde{\nu}} \left(
z/ \sqrt{3} \right) + C_2 z^{\frac{7 - c}{2}} Y_{\tilde{\nu}} \left( z/ \sqrt{3} \right), \label{ho}
\end{multline}
where $\tilde{\nu}=(- 1+ c)/2$, and $C_1$, $C_2$ are two integral constants.

Then, we try to find a particular solution for this equation. Because $z
\rightarrow 0$ rapidly after horizon-crossing, we drop out the $z^2$ term in
the coefficient of $\delta q_r$. In this case, a particular solution is given
by
\begin{equation}
\delta q_r^p \approx \frac{\gamma \Sigma_I ((12 - 3 c) \delta q_I)}{12 - 3 c} =
\gamma \Sigma_I \delta q_I. \label{par}
\end{equation}

Then we can get the general solution of Eq. \eqref{qrh}
\begin{multline}
\delta q_r = C_1 z^{\frac{7 - c}{2}} J_{\tilde{\nu}} \left(
z/ \sqrt{3} \right) + C_2 z^{\frac{7 - c}{2}} Y_{\tilde{\nu}} \left( z/ \sqrt{3} \right) + \gamma \Sigma_I \delta q_I. \label{gs}
\end{multline}

In case of $z \ll 1$, the Bessel functions in the above equation can be
approximated by
\begin{equation}
z^{\frac{7 - c}{2}} J_{\tilde{\nu}} \left(
z/ \sqrt{3} \right) \sim \frac{12^{\frac{1 - c}{4}}}{\Gamma_R \left( \frac{1 + c}{2}
\right)} z^3,
\end{equation}
\begin{multline}
z^{\frac{7 - c}{2}} Y_{\tilde{\nu}} \left(
z/ \sqrt{3} \right) \sim - \frac{1}{\pi} 12^{\frac{- 1 + c}{4}} \Gamma_R \left( \frac{- 1 + c}{2} \right ) z^{4 - c} \\
- \frac{1}{\pi} 12^{\frac{1 - c}{4}} \Gamma_R \left(
\frac{1 - c}{2} \right) \sin \left( \frac{c \pi}{2} \right) z^3,
\end{multline}
where $\Gamma_R$ is the Gamma function. From the above approximation we
know these two terms tend to zero rapidly after horizon-crossing, so
\begin{equation}
\delta q_r \approx \gamma \Sigma_I \delta q_I = - \gamma \Sigma_I \dot{\varphi_I}
\delta \varphi_I. \label{app}
\end{equation}

From Eq. \eqref{sl2} we know $\frac{4}{3} \rho_r = \gamma \Sigma_I \dot{\varphi_I}^2$, so in spatially-flat gauge the comoving curvature perturbation is given by \cite{evolution of the}
\begin{equation}
\mathcal{R} = - H \frac{\delta q}{p + \rho} = - H \frac{\Sigma_I \delta q_I + \delta
q_r}{\Sigma_I \dot{\varphi_I}^2 + \frac{4}{3} \rho_r} = - H \frac{\Sigma_I
\delta q_I}{\Sigma_I \dot{\varphi_I}^2}.
\end{equation}

Curvature perturbation $\mathcal{R}$ has a same form to that in cold inflation. In one field case, $\mathcal{R}$
reduces to our familiar form $\mathcal{R} = H \frac{\delta \varphi}{\dot{\varphi}}$.

Now, we consider a two-field model, $\varphi_1 = \phi$, $\varphi_2 =
\chi$. In this case, the perturbation equation of scalar field is given by
\begin{align}
\ddot{\delta \phi} + (3 H + \Gamma) \dot{\delta \phi} + \left( V_{\phi \phi}
+ \frac{k^2}{a^2} \right) \delta \phi + V_{\phi \chi} \delta \chi + \dot{\phi}
\delta \Gamma \nonumber \\
= \dot{\phi} \left( - \frac{k^2}{a} B + \dot{A} \right), \label{phii}\\
\ddot{\delta \chi} + (3 H + \Gamma) \dot{\delta \chi} + \left( V_{\chi \chi}
+ \frac{k^2}{a^2} \right) \delta \chi + V_{\phi \chi} \delta \phi + \dot{\chi}
\delta \Gamma \nonumber \\
= \dot{\chi} \left( - \frac{k^2}{a} B + \dot{A} \right). \label{chii}
\end{align}

Substituting Eq. \eqref{app} into Eqs.~\eqref{A} and \eqref{B}, we can express the metric perturbation $A$ and $B$ in terms of field perturbation $\delta \phi$, $\delta \chi$. In previous section, we have expressed $\delta \Gamma$ in terms of field perturbations, therefore now we get two closed differential equations for the variables $\delta \phi$, $\delta \chi$ after replacing $A$ and $B$ in above two equations with these results.

As in cold two-field inflation, we define two new adiabatic field $\sigma$
and entropy field $s$ by a rotation in field space. $d \sigma$ is tangent to
the background trajectory and $d s$ is normal to it \cite{adiabatic and entropy}.
\begin{equation}
\left( \begin{array}{c} d \sigma \\ d s \end{array} \right)=\left( \begin{array}{cc} \cos \theta \quad \sin \theta \\ -\sin \theta \quad \cos \theta \end{array} \right) \left( \begin{array}{c} d \phi \\ d \chi \end{array} \right),
\end{equation}
where $\cos \theta = \frac{\dot{\phi}}{\sqrt{\dot{\phi}^2 + \dot{\chi}^2}}$,
$\sin \theta = \frac{\dot{\chi}}{\sqrt{\dot{\phi}^2 + \dot{\chi}^2}}$

Using this definition, the equation of motion can be described in terms of
$\sigma$, $s$ is given by
\begin{equation}
\ddot{\sigma} + (3 H + \Gamma) \dot{\sigma} + V_{\sigma} = 0,
\end{equation}
\begin{equation}
\dot{\theta} \dot{\sigma} + V_s = 0,
\end{equation}
where $V_{\sigma} = \cos \theta V_{\phi} + \sin \theta V_{\chi}$, $V_s = - \sin
\theta V_{\phi} + \cos \theta V_{\chi}$.

\begin{center}
\begin{figure}[ht]
\includegraphics[width=0.45\textwidth]{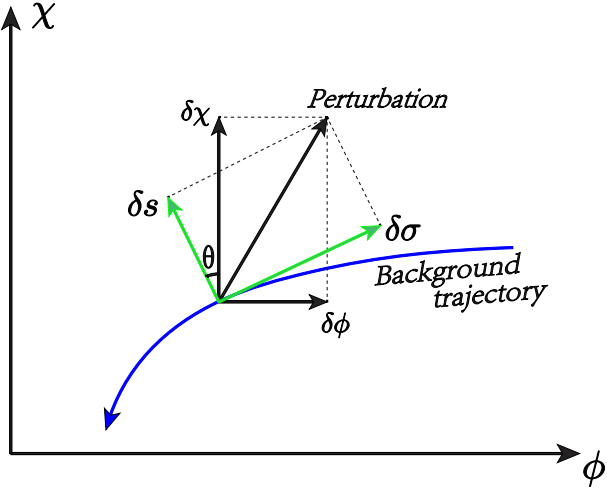}
\caption{An illustration of the decomposition of an arbitrary perturbation in field space. $\delta \phi$, $\delta \chi$ represent fluctuations with respect to a fixed local frame, and $\delta \sigma$, $\delta s$ represent fluctuations parallel and normal to the background path.}
\label{Fig.1}
\end{figure}
\end{center}

Similarly, it is useful to decompose the field perturbations into an adiabatic
$\delta \sigma$ and entropy $\delta s$ component as illustrated in Fig.~\ref{Fig.1}, $\delta \sigma$ is parallel to the background trajectory and $\delta
s$ is orthogonal to it.
\begin{equation}
\left( \begin{array}{c} \delta \sigma \\ \delta s \end{array} \right)=\left( \begin{array}{cc} \cos \theta \quad \sin \theta \\ -\sin \theta \quad \cos \theta \end{array} \right) \left( \begin{array}{c} \delta \phi \\ \delta \chi \end{array} \right).
\end{equation}

Now we can write Eqs.~\eqref{phii} and \eqref{chii} in the form
\begin{multline}
\ddot{\delta \sigma} + (3 H + \Gamma) \dot{\delta \sigma} + \left(
\frac{k^2}{a^2} + V_{\sigma \sigma} - \dot{\theta}^2 \right) \delta \sigma - 2 \dot{\theta} \dot{\delta s} \\
 + 2 \left( \frac{2 \dot{\theta}
V_{\sigma}}{\dot{\sigma}} - \ddot{\theta} \right) \delta s + \dot{\sigma}
\delta \Gamma \\
 = - \frac{k^2}{a} \dot{\sigma} B - (\dot{\sigma} \Gamma + 2
V_{\sigma}) A + \dot{\sigma} \dot{A}, \label{sigma}
\end{multline}
\begin{multline}
\ddot{\delta s} + (3 H + \Gamma) \dot{\delta s} + \left( \frac{k^2}{a^2} +
V_{ss} - \dot{\theta}^2 \right) \delta s \\
+2 \dot{\theta} \dot{\delta \sigma}
- \frac{2 \dot{\theta} \ddot{\sigma}}{\dot{\sigma}} \delta \sigma = 2
\dot{\theta} \dot{\sigma} A, \label{s}
\end{multline}
where
\begin{eqnarray}
V_{\sigma \sigma} &=& \cos^2 \theta V_{\phi \phi} + \sin 2 \theta V_{\phi \chi} + \sin^2 \theta V_{\chi \chi}, \\
V_{s s} &=& \sin^2 \theta V_{\phi \phi} - \sin 2 \theta V_{\phi \chi} + \cos^2 \theta V_{\chi \chi}.
\end{eqnarray}

Since $\Gamma = \Gamma (\phi, \chi, \rho_r)$ in previous section, now we can
treat $\Gamma$ as $\Gamma = \Gamma (\sigma, s, \rho_r)$. For simplicity, we
redefine some slow-roll parameters
\begin{gather}
\epsilon_{\sigma} = \frac{\dot{\sigma}^2}{2 H^2}, \quad \eta_{\sigma} = -
\frac{\ddot{\sigma}}{H \dot{\sigma}}, \quad \eta_r = - \frac{1}{2}
\frac{\dot{\rho_r}}{H \rho_r}, \nonumber \\
\beta_{\sigma} = \frac{V_{\sigma} \Gamma_{\sigma}}{V \Gamma}, \quad \beta_s = \frac{V_{\sigma} \Gamma_s}{V \Gamma}.
\end{gather}
Then, the metric perturbations $A$, $B$ in Eqs.~\eqref{A}, \eqref{B} and $\delta \Gamma = \Gamma_{\sigma} \delta \sigma + \Gamma_s \delta s + \Gamma_{\rho_r} \delta \rho_r$ can be expressed in term of $\delta \sigma, \delta s$ (According to Eqs.~\eqref{momentum}, \eqref{A} and \eqref{app}, we can rewrite $\delta \rho_r$ in terms of $\delta \sigma,
\delta s$ in spatially-flat gauge).

Substituting $\delta \Gamma, A, B$ into Eqs.~\eqref{sigma} and \eqref{s}, we get
\begin{multline}
\ddot{\delta \sigma} + \left( 3 H + \Gamma + \frac{\Gamma \dot{\sigma}^2}{3
H^2} + \frac{\Gamma \dot{\sigma}^2 \Gamma_{\rho_r}}{H} \right) \dot{\delta
\sigma} + \left ( \frac{k^2}{a^2}- \dot{\theta}^2 - 3 \dot{\sigma}^2 - \right. \\
\left. \frac{3 \Gamma \dot{\sigma}^2}{2 H} \! -\! \frac{\Gamma^2 \dot{\sigma}^2}{6 H^2} \!+\! V_{\sigma \sigma}\! -\! \frac{\dot{H} \Gamma \dot{\sigma}^2 \Gamma_{\rho_r}}{H^2} \!-\! \frac{2 \rho_r \dot{\sigma}^2 \Gamma_{\rho_r}}{H}\! -\! \frac{2 \Gamma \rho_r \dot{\sigma}^2 \Gamma_{\rho_r}}{3 H^2} \right. \\
\left. + \frac{\Gamma \dot{\sigma} \ddot{\sigma} \Gamma_{\rho_r}}{H} + \frac{\dot{\rho_r} \dot{\sigma}^2 \Gamma_{\rho_r}^2}{H} + \dot{\sigma} \Gamma_{\sigma} + \frac{\dot{\sigma}^3 \Gamma_{\sigma}}{3 H^2} + \frac{\dot{\sigma}^3 \Gamma_{\rho_r} \Gamma_{\sigma}}{H} \right ) \delta \sigma \\
  = 2 \dot{\theta} \dot{\delta s} + \left( \frac{\dot{\theta} \dot{\sigma}^2}{H} - \frac{2 \dot{\theta} V_{\sigma}}{\dot{\sigma}} + 2 \ddot{\theta} - \dot{\sigma} \Gamma_s \right) \delta s, \label{sigma2}
 \end{multline}
 \begin{multline}
\ddot{\delta s} + (3 H + \Gamma) \dot{\delta s} + \left( \frac{k^2}{a^2} +
V_{ss} - \dot{\theta}^2 \right) \delta s \\
= - 2 \dot{\theta} \dot{\delta
\sigma} + \left( \frac{(3 H + \Gamma) \dot{\sigma}^2}{3 H^2} + \frac{2
\ddot{\sigma}}{\dot{\sigma}} \right) \dot{\theta} \delta \sigma. \label{s2}
\end{multline}

The comoving curvature perturbation is given by
\begin{equation}
\mathcal{R} = H \frac{\dot{\phi} \delta \phi + \dot{\chi} \delta \chi}{\dot{\phi}^2 + \dot{\chi}^2} = H \frac{\delta \sigma}{\dot{\sigma}}. \label{ccp}
\end{equation}
Replacing $\delta \sigma$ in Eq. \eqref{sigma2} with $\mathcal{R}$, we can rewrite Eqs.~\eqref{sigma2} and \eqref{s2} as two coupled differential equations of $\mathcal{R}$, $\delta s$. If we keep only the leading order in the slow-roll approximation, the equations are given by
\begin{multline}
\ddot{\mathcal{R}} + 3 (1 + \gamma + c \gamma) H \dot{\mathcal{R}} +\left ( \frac{k^2}{a^2} -
\dot{\theta}^2 \right ) \mathcal{R} \\
= \frac{2 H \dot{\theta}}{\dot{\sigma}} \dot{\delta s}
+ \left ( \frac{2 H \dot{\theta}}{\dot{\sigma}} (3 + 3 \gamma) H + \frac{2 H \ddot{\theta}}{\dot{\sigma}} \right ) \delta s, \label{rr}
\end{multline}
\begin{equation}
\ddot{\delta s} + 3 (1 + \gamma) H \dot{\delta s} + \left( \frac{k^2}{a^2} + M_{\rm eff}^2 \right) \delta s = -
\frac{2 \dot{\theta} \dot{\sigma}}{H} \dot{\mathcal{R}}, \label{ss}
\end{equation}
where $M_{\rm eff}^2=V_{s s}- \dot{\theta}^2$, $M_{\rm eff}$ is the effective mass of $\delta s$.

According to Eqs.~\eqref{rr} and \eqref{ss}, we know when we neglect the curvature of background trajectory in field space ($\dot{\theta} = 0$), $\delta s$ behave like a free field, and when $r=0$, these equations can go back to cold inflation.

Now we define the isocurvature perturbation $\mathcal{S} = \frac{H}{\dot{\sigma}} \delta s$ \cite{spectra running and},
and the power spectrum of curvature and isocurvature perturbation \cite{on the importance}
\begin{eqnarray}
\mathcal{P}_{\mathcal{R}} &=& \frac{k^3}{2 \pi^2} \left| \mathcal{R} \right|^2, \\
\mathcal{P}_{\mathcal{S}} &=& \frac{k^3}{2 \pi^2} \left| \mathcal{S} \right|^2.
\end{eqnarray}
The spectral index of the curvature perturbation is
\begin{equation}
n_s-1 = \frac{d \ln \mathcal{P}_{\mathcal{R}}}{d \ln k}.
\end{equation}

The tensor modes of perturbations are not affected by the thermal noises, so the tensor power spectrum and tensor-to-scalar ratio at the pivot scale are given by \cite{inflation and the}
\begin{eqnarray}
\mathcal{P}_T &=& 8 \left( \frac{H_*}{2 \pi} \right)^2, \\
r &=& \frac{\mathcal{P}_{\mathcal{R}}}{\mathcal{P}_T}.
\end{eqnarray}

In this work, we are mainly concerned with the large scale evolution of curvature perturbation $\mathcal{R}$, because the value of $\mathcal{R}$ at the end of inflation seeds the observed CMB temperature anisotropies, corresponding to the variance of inhomogeneities' distribution.

\section{\label{se4}Numerical Examples and Constraints from Observations}

\subsection{\label{se4.1} Numerical Examples}

When dealing with multi-component systems, a numerical method is almost essential. In this section we use the formalism introduced above to investigate a toy model, in which massive scalar fields $\phi$ and $\chi$ are coupled through an interaction term $\frac{1}{2} g^2 \phi^2 \chi^2$ \cite{correlation-consistency cartography of}.
\begin{equation}
V (\phi, \chi) = \frac{1}{2} m_{\phi}^2 \phi^2 + \frac{1}{2} m_{\chi}^2
\chi^2 + \frac{1}{2} g^2 \phi^2 \chi^2. \label{vv}
\end{equation}

The background equations are:
\begin{gather}
\ddot{\phi}+(3H + \Gamma)\dot{\phi} + \frac{\partial V(\phi,\chi)}{\partial \varphi}=0 \label{phiphi} \\
\ddot{\chi}+(3H + \Gamma)\dot{\chi} +\frac{\partial V(\phi,\chi)}{\partial \chi} = 0 \\
\dot{\rho_r}+4 H \rho_r = \Gamma (\dot{\phi}^2 + \dot{\chi}^2)\\
\qquad H^2 = \frac{1}{3} \left ( \frac{1}{2} \dot{\phi}^2 + \frac{1}{2} \dot{\chi}^2 + V(\phi,\chi) + \rho_r \right ) \label{HH}
\end{gather}

There are five free parameters associated with the initial conditions of the
equation of motion, $\phi_0$, $\chi_0$, $\dot{\phi_0}$, $\dot{\chi_0}$, $\rho_{r 0}$. After making use of slow-roll approximation, $\dot{\phi} = - \frac{1}{3 H + \Gamma} \frac{\partial V(\phi,\chi)}{\partial \phi}$,
$\dot{\chi} = - \frac{1}{3 H + \Gamma} \frac{\partial V(\phi,\chi)}{\partial \chi}$, $\rho_r = \frac {\Gamma}{4H} (\dot {\phi}^2 + \dot {\chi}^2)$, the initial conditions
are given by $\phi_0$, $\chi_0$. We choose the parameters associated with
potential to be $m_{\phi} = 2 \times 10^{- 7}$, $m_{\chi} = 10^{- 6}$, $g = 2
\times 10^{- 8}$, and give a numerical result below.

In our numerical calculations, we set $\gamma = \frac{\Gamma}{3 H}$ to be a constant. In order to get a clear picture of the evolution of background and
perturbation variables, we integrate the exact background equations \eqref{phiphi}-\eqref{HH} first until horizon-crossing. After horizon-crossing, we integrate the background and perturbation equations \eqref{rr}-\eqref{ss} simultaneously to the end of inflation. We set the effective particle number of radiation $g_{\ast} = 228.75$ \cite{observational constraints on}, and choose the initial values of curvature perturbation $\mathcal{R}$ and entropy perturbation $\delta s$ at horizon-crossing according to Eq. \eqref{ip}. In addition, we take the number of e-foldings from horizon-crossing to the end of inflation to be $\Delta N =60$ to make definite calculations.

According to the top left panel of Fig.~\ref{Fig.2} we know that the heavy field $\chi$ decrease faster than the light field $\phi$, and after a period of time, $\chi$ reaches zero and then inflation will be driven by one single field $\phi$. the bottom two graphs of Fig.~\ref{Fig.2} show that the potential dominates the total energy in inflationary period, which is consistent with slow-roll condition. However, radiation density will increase rapidly at the end of inflation and then become dominated, at the same time slow-roll conditions break down.

The left panel of Fig.~\ref{Fig.3} shows after horizon-crossing $\frac{k}{a H} \rightarrow 0$ quickly and $\mathcal{R}$, $\delta s$ tend to a constant value and they are weakly correlated. From Eqs.~\eqref{rr} and \eqref{ss} we know $\dot{\theta}$ plays an important role in the interaction of curvature perturbation $\mathcal{R}$ and entropy perturbation $\delta s$. The upper left panel of Fig.~\ref{Fig.2} shows at around $N = 80$ the heavy field $\chi$ decays to zero and this will cause a bump in $\dot{\theta}$. At around $N = 80$, $\dot{\theta}$ increases suddenly and there is a strong interaction between $\mathcal{R}$ and $\delta s$, and they all change significantly. After that, the entropy perturbation decay to zero and the curvature perturbation $\mathcal{R}$ become a nearly constant value again.

\begin{widetext}
\begin{center}
\begin{figure}[ht]
\begin{tabular}{cc}
\includegraphics[width=0.85\textwidth]{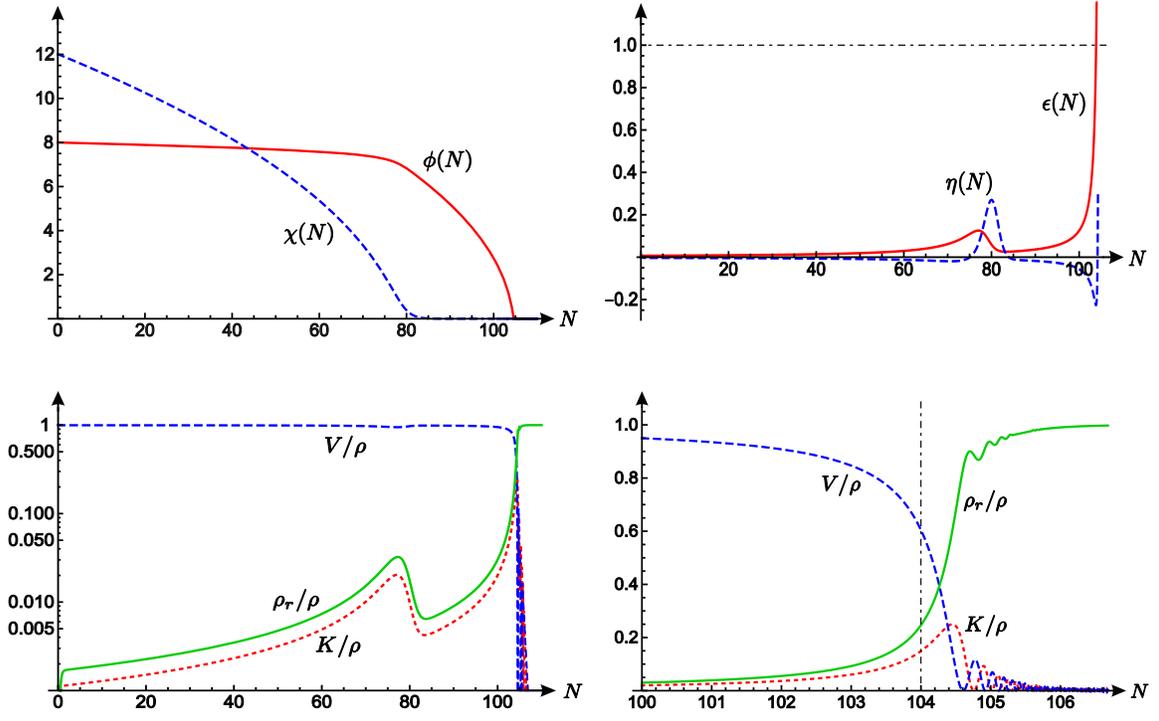}
\end{tabular}
\caption{The evolution of different background variables are shown against e-foldings $N$($d N = H d t$) in case of $r = 1$, $\phi_0 = 8, \chi_0 = 12$, and we have set $N_0 = 0$ at the initial time. Top left and top right panel: the evolution of scalar fields and the slow-roll parameter $\epsilon$, $\eta$. Bottom left and bottom right panel show the evolution of kinetic energy density $K$ ($K = \frac{1}{2}
\dot{\phi}^2 + \frac{1}{2} \dot{\chi}^2$), potential $V$ and radiation energy density
$\rho_r$, and all the energy is scaled by total energy density $\rho$ in the
graph. In bottom right panel, we zoom in to the few e-foldings around the end of inflation to give more details, and the dot-dashed vertical line indicates the end of inflation ($\epsilon = 1$). In this case inflation ends at about $N=104$.}
\label{Fig.2}
\end{figure}
\end{center}
\end{widetext}

\begin{widetext}
\begin{center}
\begin{figure}[ht]
\begin{tabular}{cc}
\includegraphics[width=0.85\textwidth]{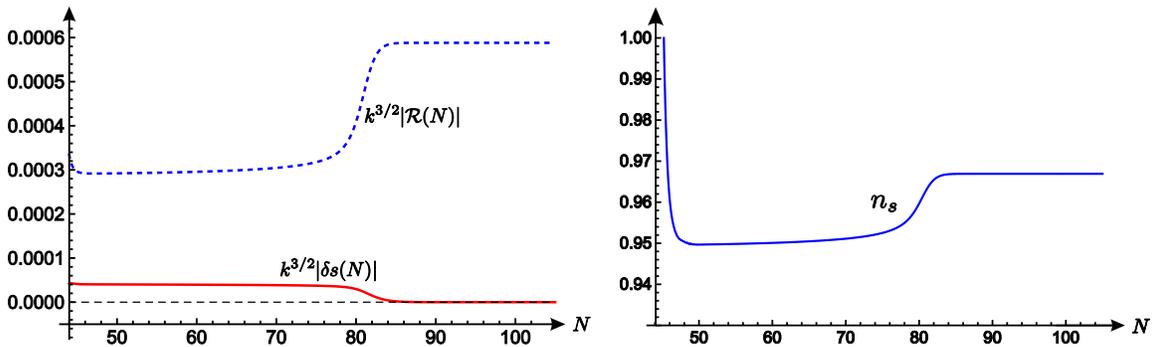}
\end{tabular}
\caption{The evolution of curvature perturbation $\mathcal{R}$, entropy perturbation $\delta s$ and spectral index $n_s$ in $\Gamma \propto H$ case. The initial value of $\mathcal{R}$, $\delta s$ are chosen based on Eq. \eqref{ip}. The spectral index $n_s$ is calculated using finite-difference method.}
\label{Fig.3}
\end{figure}
\end{center}
\end{widetext}

\subsection{\label{se4.2} Constraints from Observations}

The most recent measurements of the cosmic microwave background (CMB) provides narrow constraints on cosmological parameters, ruling out large classes of models. Having established representative examples in previous subsection, we now turn our attention to the compatibility with observational data. We will use the most recent Plank data to constrain our models, finding the allowed regions of parameter space consistent with the observational values of $n_s$ and $r$. For simplicity, we neglect the interaction between $\phi$ and $\chi$ (set $g=0$), and introduce the mass ratio defined as $R_m = m_{\chi} / m_{\phi}$.
When we fix the value of $R_m$, for a given $(\gamma, m_{\phi})$ when $\Gamma \propto H$, every set of initial condition $(\phi_{\ast}, \chi_{\ast})$ will produce a corresponding e-folding $N$, and a corresponding curvature power spectrum $\mathcal{P}_{\mathcal{R}}$
at the end of inflation ($\epsilon = 1$), just like the numerical examples shown in last subsection. Then we use the condition $N = 60$ and $\mathcal{P}_{\mathcal{R}} = 2.207 \times 10^{- 9}$(68\%CL, {\emph{Plank}}
TT,TE,EE+LowP) to constrain the parameter space and pick out the exact initial condition $(\phi_{\ast}, \chi_{\ast})$ for each value of $(\gamma, m_{\phi})$. Next we give a numerical result of $(r,n_s)$ at the end of inflation with the initial condition we obtained. That is to say, we have a set of $(r, n_s)$ for every $(\gamma, m_{\phi})$, so we can determine the range of our parameters in face of the observational results of $(r, n_s)$.

We use the observational data $n_s = 0.9645 \pm 0.0049$, $r < 0.1$ (68\% CL, {\emph{Planck}} TT,TE,EE+LowP) to constrain our models in case of $R_m = 3$ and $R_m=10$. Our results are given in parameter space of $(\gamma, m_{\phi})$ and $(r, n_s)$ plotted in Fig.~\ref{Fig.4.eps}. First, we neglect the points (in white areas) which do not have a corresponding initial condition of $(\phi_{\ast}, \chi_{\ast})$ to produce expected e-foldings and power spectrum at the same time. Then we proceed to search for the observationally allowed regions of parameters from the remaining points. In the plot, the light-gray shaded regions indicate areas for $r > 0.1$ while the dark-gray shaded regions indicate areas for $r < 0.1$. The regions highlighted in red are for $0.9596 < n_s < 0.9694$, so the intersections of dark-gray shaded areas and red shaded areas give the regions of parameter space consistent with the observational data.
\begin{widetext}
\begin{center}
\begin{figure}[ht]
\begin{tabular}{cc}
\includegraphics[width=0.85\textwidth]{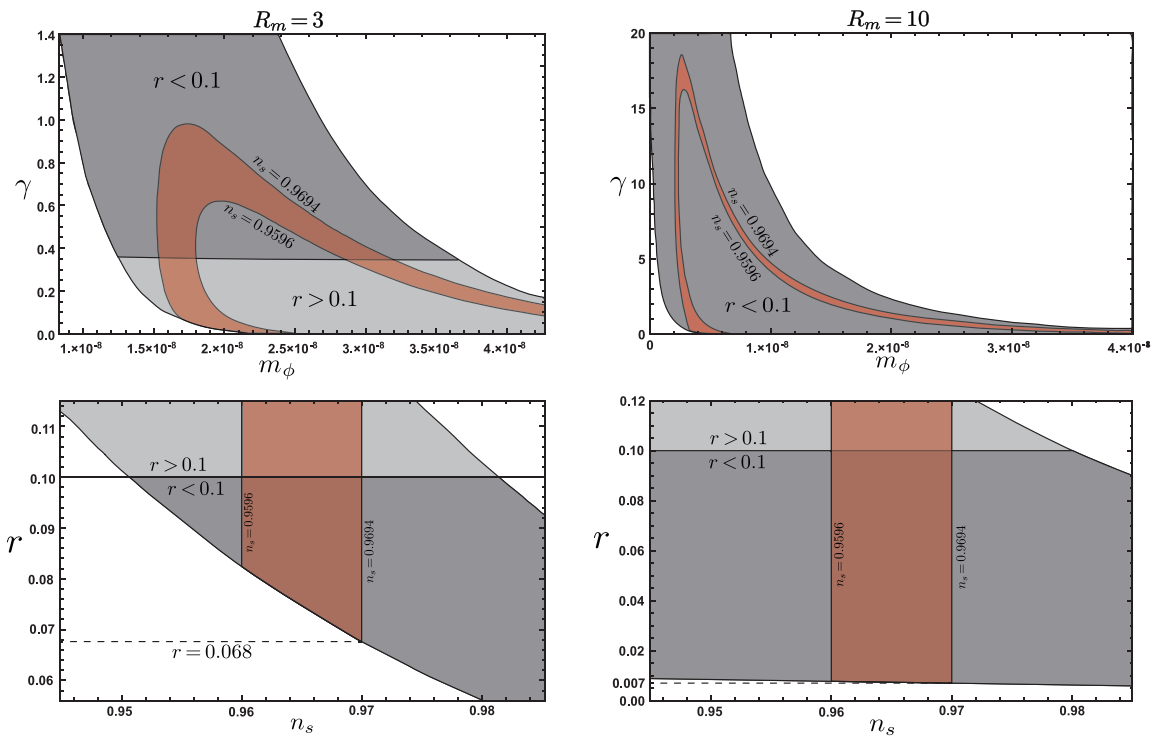}
\end{tabular}
\caption{In the above $(\gamma, m_{\phi})$ or $(r, n_s)$ planes, we show constraints of observational data on the spectral index $n_s$ and tensor-to-scalar ratio $r$. The intersections of dark-gray shaded areas and red shaded areas give the observationally allowed regions. In the left two panels we take the mass ratio $R_m=3$, and in the right two panels we take $R_m=10$. For each value of $R_m$, results are shown in both $(\gamma,m_{\phi})$ planes and $(r,n_s)$ planes.}
\label{Fig.4.eps}
\end{figure}
\end{center}
\end{widetext}

According to Fig.~\ref{Fig.4.eps}, we know all models in our analysis have some observational allowed regions in parameter plane. As the plots show, thermal fluctuations are much stronger than quantum fluctuations in warm inflation, so we need smaller masses of the scalar field to produce observationally allowed power spectrum. In upper two panels, we show when $\Gamma \propto H$, different $R_m$ leads to different permitted range of dissipative strength $\gamma$, and inflation can happen in both weak and strong regime of warm inflation. In $R_m = 3$ case, $\gamma$ takes values of $0.38 \lesssim \gamma \lesssim 1.0$, and when $R_m = 10$, $0.4 \lesssim \gamma \lesssim 19$. The lower two panels give the lower bound of tensor-to-scalar ratio $r$ in the observationally allowed range of spectral index $n_s$, depending on the value of $R_m$. As illustrated, observational data favour large value of $R_m$ in two-field cases. For $R_m=3$, $r$ gets a lower bound $r \approx 0.068$, this is a rather large value and may become disfavored by the observations in the near future. In case of $R_m=10$, we obtain a much smaller bound of $r \approx 0.007$, which is in good agreement with the observational constraints.
\section{\label{se5}Conclusions}

In this paper, we have studied inflation driven by multiple scalar fields and
an interacting perfect fluid. We defined some new parameters and perform a full analysis of perturbation equations, including field perturbation, fluid perturbation and metric perturbation.  Then we apply the theory to warm inflation, and give the evolution equations of curvature perturbation $\mathcal{R}$ and isocurvature perturbation $\delta s$ in a two-field case. Next, we perform numerical calculations in our representative examples and give the main features of the evolution of background and perturbation variables. Finally, in order to check the compatibility of our models with observations, we use the most up-to-date observational data to constrain our model and give the observational permitted regions of the parameters. In the calculation of perturbations, we have used the slow-roll approximation for simplicity.

According to the numerical results above, the correlation between curvature and entropy perturbations can change $\mathcal{P}_\mathcal{R}$ significantly on large scales, and the change mainly occurs simultaneously to the turning of the background trajectory. However, the change cannot be observed if it happens much more than 60 e-foldings before the end of inflation \cite{computing observables in}, in which case the effects of multi-field are negligible. The damping term in background equation can slow the decrease of scalar fields, so warm inflation will produce more e-foldings after the turning of trajectory. Consequently, the multi-field effect is more likely to be observed when the damping effect is not too strong. Fortunately, the Planck satellite has put a tight upper bound on the primordial non-Gaussianity, and warm inflationary models tend to produce large non-Gaussianity, so the observations seems not to be compatible with the very strong version of warm inflation \cite{non-gaussianity in fluctuations}.

Our results also show that the inflaton field may starts to oscillate after the end of inflation, just as the reheating phase in cold inflation. This is a common phenomenon in weak regime of warm inflation, and different forms of the damping coefficient can lead to different dynamical features of warm inflation. In fact, many models lie between warm inflation and standard inflation, and there may be some general framework to describe them \cite{extended warm inflation}. The existence of radiation will not alter the main features of super-horizon evolution of scalar perturbations during slow-roll regime, compared with standard inflation, which is consistent with observations. However, the value of comoving curvature perturbation at the end of inflation depends on the form of damping coefficient $\Gamma$.

To study the compatibility with observations, we compared the predictions of our models with the most up-to data observational data, and show the results in Fig.~\ref{Fig.4.eps}. In all the cases considered, we find some observationally allowed regions in parameter space and the mass ration $R_m$ has a significant impact on the allow regions of parameters. As illustrated by Fig.~\ref{Fig.4.eps}, the models with a smaller value of $R_m$ tend to be more constrained by observational data. In our results, the observationally permitted range of the mass of the lighter field $m_{\phi}$ are all less than $5 \times 10^{- 8} M_p$ ($M_p$ is the reduced Planck mass), which is much smaller than the double inflation models in cold inflation \cite{double inflation and}. This is easy to understand because in warm inflation thermal fluctuations are much stronger than quantum fluctuation, and we do not need large $m_{\phi}$ to produce the expected value of scalar power spectrum. Note that if we set $\gamma = 0$ in our models, we go back to the cold inflation regime. However, according to upper two panels Fig.~\ref{Fig.4.eps} we know this case is not observational allowed for it predicts too large scalar-to-tensor ratio. Therefore, we can conclude that warm inflation effects can reduce the value of $r$ in multi-field cases, making more inflationary models fit the observations.

In our numerical analysis, we take a $T$-independent dissipative coefficient as an example, which may not be a realistic case \cite{warm little inflation}. As shown in \cite{density fluctuations from}, the $T$-dependent dissipative coefficient leads to a growing mode in the fluctuations before horizon-crossing in case of $\gamma > 1$ \cite{cosmological fluctuations of,shear viscous effects}. This is an important effect in warm inflation and has to be taken into account. In $T$-dependent cases the perturbations need to be computed numerically, and we leave this for our future work. The effective mass of isocurvature mode $M_{\rm eff}$ in the models we studied here is large enough so that the isocurvature perturbation will decay to zero before the end of inflation. However, this is not always the case \cite{isocurvature perturbations and}, and further research should be done on this topic. Besides, when dealing with $\mathcal{N} \gg 1$ scalar fields in inflationary models, the method of constructing random potentials is worth considering \cite{charting an inflationary}.

\section{Acknowledgements}
This work was supported by the National Natural Science Foundation of China (Grants No. 11575270, No. 11175019, No. 11235003 and No.11605100).

\end{document}